# Bulk and surface characterization of $In_2O_3$(001) single crystals

Daniel R. Hagleitner,[1] Peter Jacobson,[1] Sara Blomberg,[2] Karina Schulte,[2] Edvin Lundgren,[2] Markus Kubicek,[3] Jürgen Fleig,[3] Frank Kubel,[3] Christoph Puls,[3] Andreas Limbeck,[3] Herbert Hutter,[3] Lynn A. Boatner,[4] Michael Schmid,[1] Ulrike Diebold[1]

[1] Institute of Applied Physics, TU Wien, Wiedner Hauptstr. 8-10/134, 1040 Wien, Austria
[2] Division of Synchrotron Radiation Research, Lund University, Box 118, SE-22100, Sweden
[3] Institute of Chemical Technologies and Analytics, TU Wien, Getreidemarkt 9/164, 1060 Wien, Austria
[4] Materials Science and Technology Division, Oak Ridge National Laboratory, Oak Ridge, Tennessee 37831

A comprehensive bulk and surface investigation of high-quality $In_2O_3$(001) single crystals is reported. The transparent-yellow, cube-shaped single crystals were grown using the flux method. ICP-MS measurements reveal small residues of Pb, Mg and Pt in the crystals. Four-point-probe measurements show a resistivity of $2 \pm 0.5 \times 10^5$ Ω cm, which translates into a carrier concentration of $\lesssim 10^{12}$ $cm^{-3}$. The results from X-ray diffraction (XRD) measurements revise the lattice constant to 10.1150(5) Å from the previously accepted value of 10.117 Å. Scanning Tunneling Microscopy (STM) images of a reduced (sputtered/annealed) surface show a step height of 5 Å, which indicates a preference for one type of surface termination. A combination of low-energy ion scattering (LEIS) and atomically resolved STM indicates an indium-terminated surface with small islands of 2.5 Å height, which corresponds to a strongly distorted indium lattice. Scanning Tunneling Spectroscopy (STS) reveals a pronounced surface state at the Fermi Level ($E_F$). Photoelectron Spectroscopy (PES) shows additional, deep-lying band gap states, which can be removed by exposure of the surface to activated oxygen. Oxidation also results in a shoulder at the O 1s core level at a higher binding energy, possibly indicative of a surface peroxide species. A downward band bending of 0.4 eV and an upward band bending of ~0.1 eV is observed for the reduced and oxidized surfaces, respectively.

## 1 Introduction

Indium oxide, $In_2O_3$, has drawn increased attention from researchers over recent years. When doped with $SnO_2$, the material is commonly referred to as Indium Tin Oxide (ITO), which is the prototypical Transparent Conducting Oxide (TCO). ITO combines high optical transparency in the visible range with conductivity approaching that of a metal.[1] As a consequence of its interesting physical and chemical characteristics, it is widely used in several technical applications including transparent films for Organic Light Emitting Diodes (OLEDS) and organic photovoltaic cells (OPVC), gas sensing, and transparent infrared reflectors. $In_2O_3$ is also used in heterogeneous catalysis[2] and in chemical gas sensing.[3] Despite its technological importance, surprisingly little is known about the fundamental surface properties of ITO or those of pure $In_2O_3$. Even basic material characteristics, such as the fundamental band gap, are the subject of some continuing controversy.[4,5] The use of a value of 3.75 eV[6] for $In_2O_3$ (which is the optically-determined band gap) instead of the actual value of around 2.9 eV[4] has led to many incorrect conclusions - especially regarding band bending effects and defect chemistry.

Another problem associated with previous investigations is that the samples were produced mainly by magnetron sputtering, evaporation, or by pressing pellets of powder, which usually leads to polycrystalline, highly defective samples[7] - although new information could be gained from epitaxially-grown $In_2O_3$ and ITO thin films.[4,8-10] In the present work, high quality, single crystals of $In_2O_3$ are examined.

$In_2O_3$ is an n-type semiconductor. The n-type conduction arises from intrinsic donor defects, which also explains the pronounced non-stoichiometry observed under highly reducing conditions.[11] So far, $In_2O_3$ was regarded as a very good conductor even without extrinsic dopants due to its high carrier concentration.[12]

$In_2O_3$ can exist in three different modifications. In this work, only single-crystal samples crystallized in the bixbyite structure (space group $Ia\overline{3}$) are utilized. The bixbyite structure can be derived from a 2×2×2 fluorite lattice.[13] To obtain the bixbyite structure, 12 oxygen atoms per unit cell are removed from the anion sublattice in a systematic way. Each {001} anion layer contains four of these vacant oxygen positions. The cubic unit cell of $In_2O_3$ has – according to the literature[14] – a lattice parameter of 10.117 Å (for a revised value see Sec. 2.1.2) and consists of 80 atoms; 32 indium atoms and 48 oxygen atoms (Figure 1). Two distinct indium positions can be found in the crystal, that have different symmetries. These are commonly referred to as In-b (8 atoms) and In-d (24 atoms). The In-b sites are more regularly coordinated than the In-d sites. Along the (001) direction, the crystal can be thought of as a stacking of three different layers. These can be identified as a mixed layer, which consists of In-b and In-d atoms (M-layer), a layer which contains only In-d atoms (D-layer), and an oxygen layer. An important characteristic of $In_2O_3$(001) is the polarity of its surface (Tasker type-III)[15,16]: The alternating indium and oxygen layers lead to a net dipole moment perpendicular to the surface. However, polar surfaces cannot have bulk-like surface terminations: in this case the electrostatic potential would diverge for a macroscopic material. A depolarization field is, therefore, required to stabilize the surface. This can be achieved by different mechanisms such as reducing the top and bottom layer surface charges, changes in surface stoichiometry, surface reconstructions, absorption or faceting.[16] Recent STM measurements of epitaxial ITO films showed that the non-polar (111) surface has a (1x1) structure with a simple bulk-like termination[8], while the polar (001) surface is considerably more complex.[9] These previous measurements were, however, complicated by the fact that (001)-oriented $In_2O_3$ films tend to facet, which leads to rough surfaces.[10] The availability of high-quality, (001)-oriented single crystals, in combination with a recent DFT work[17] that describes possible terminations of low-index $In_2O_3$ surfaces, has motivated us to undertake the present study.

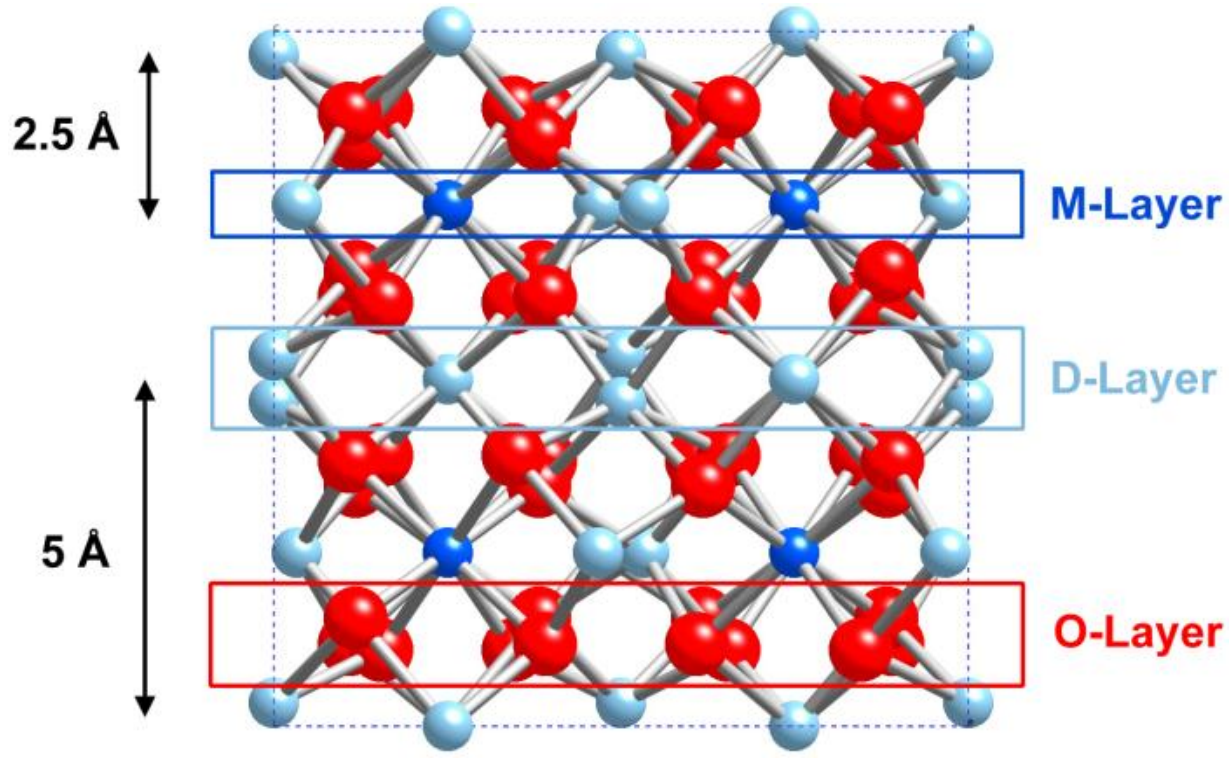

Figure 1 (Color online) Unit cell of the $In_2O_3$ bixbyite structure. Large (red) balls oxygen, small (dark) balls In-b and In-d (bright).

# 2 Experiments

## *2.1 Bulk Characteristics*

### 2.1.1 Crystal Growth

The crystals were grown using a flux method. The flux (solvent) used was $B_2O_3$ and PbO. In addition, a small amount of MgO was added. This mixture of $In_2O_3$, $B_2O_3$, PbO and MgO powders was melted in a platinum crucible and maintained at a temperature of 1200 °C for 4 to 10 hours. The temperature was then programmed to decrease by 3 °C per hour. As soon as a temperature of 500°C was reached, the furnace was turned off. The crystals were extracted from the solidified flux using a 1:4 solution of nitride acid and water.

The resulting cubic crystals are yellow-transparent and have side lengths between 1 and 2 mm as show in Figure 2. Polarized light microscopy showed that the crystals can be characterized as heterogeneous single crystals with multiple growth domains and stress birefringence (see Figure 2).

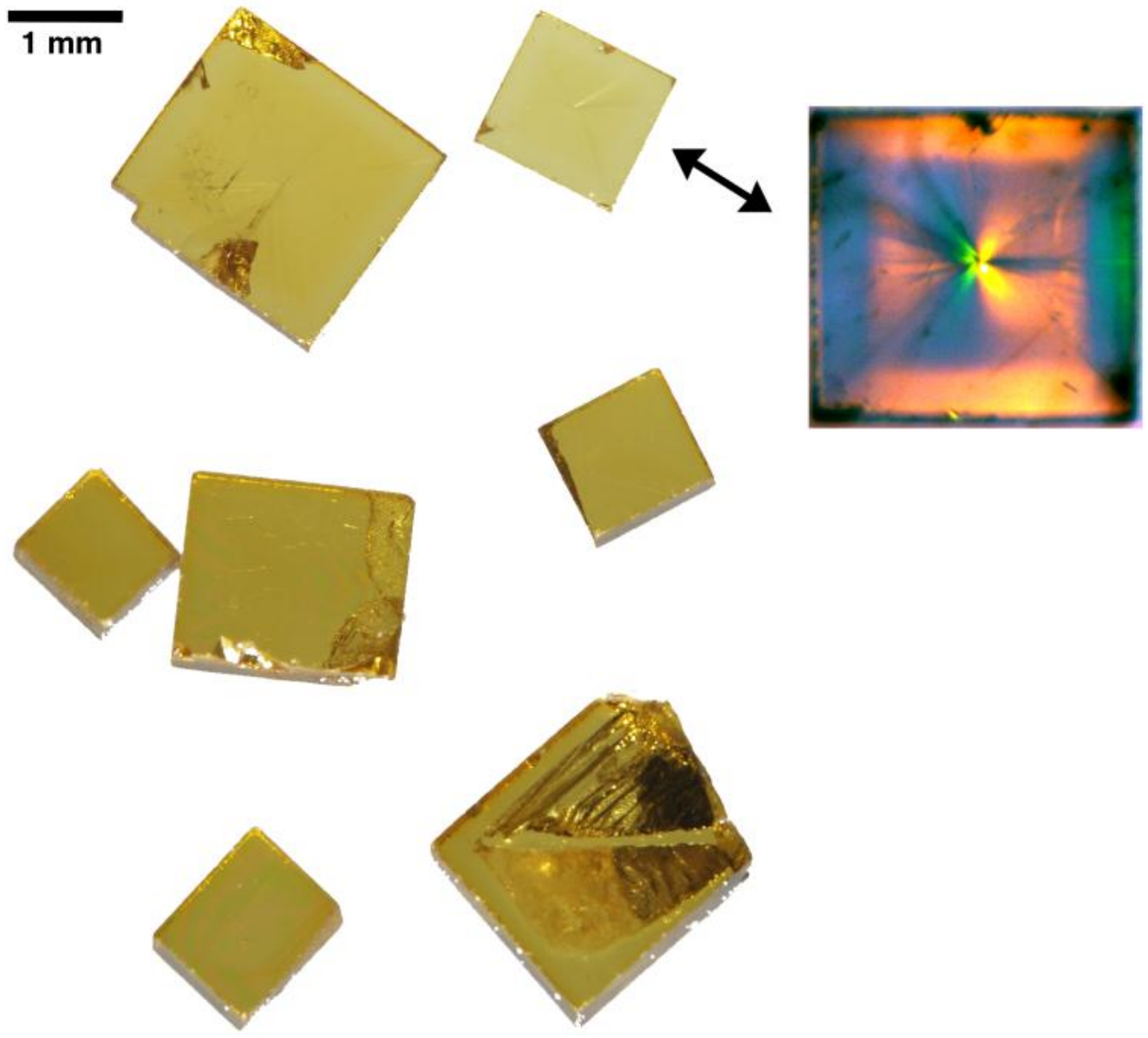

Figure 2 (Color online) The investigated $In_2O_3$ single crystals. A polarized light microscope image of one of the crystal is shown.

### 2.1.2 ICP-MS/SIMS

ICP measurements were performed to determine whether the crystals have residues from the flux or show other unknown contaminations. The measurements were performed using an NWR-213 laser ablation system from New Wave Research (ESI, Fremont, CA) coupled to a X-Series 2 ICP-MS System from Thermo Fisher Scientific (Bremen, Germany). A total of 18 different spots were ablated to minimize the influence of local irregularities in the sample. Trace impurities were quantified using a mass balance model based on the assumption that the target is 100 % $In_2O_3$ with a natural isotope distribution. Element concentrations were calculated from the acquired intensity counts with reference to the $^{113}In$ signal and the respective natural isotope abundance. Results were averaged from all 18 measurements. The highest concentrations were found for Pb (4307 ± 74 ppm), Mg (1388 ± 42 ppm) and Pt (155 ± 28 ppm). Trace amounts (< 50 ppm) of Zr, Sn, Sb, Nd and Bi were identified. Boron has a low signal intensity in ICP-MS and only a small amount was detected.

The TOF-SIMS measurements were carried out on a TOFSIMS.5 (ION-TOF GmbH, Münster, Germany) instrument. Depth profiles (~2000 nm) for the most abundant impurities in the crystals (Pb, Mg, Pt) showed a homogenous distribution and no significant reduction of intensity with sputter time.

### 2.1.3 XRD

A small fragment broken from the corner region of the platelet was used for single crystal analysis. This crystal was controlled optically and showed no birefringence or domains. Structure analysis was made using a Bruker

SMART diffractometer with a graphite monochromator and a SMART APEX detector using molybdenum $K_\alpha$ radiation ($\lambda$ = 0.71073 Å) at room temperature (293 K). Frames were integrated using SAINT PLUS (Bruker, 2008) and absorption correction was performed using a multi-scan approach with SADABS (Bruker, 2008). The lattice parameters and systematic extinctions clearly indicate an orthorhombic space group $Ia\overline{3}$. The unit cell size was determined to 10.1150(5) Å. The structure analysis was made using the XTAL refinement package and converged to R=0.017.[18] The atomic coordinates are given in Table 1. Details of the refinement are given in the Table A.1, atomic displacement parameters can be found in Table A.2 and inter-atomic distances in Table A.3.

| Site | Wyckoff Symbol | x | Y | Z |
|---|---|---|---|---|
| In1 | 8b | 0.25 | 0.25 | 0.25 |
| In2 | 24d | 0.466393(2) | 0 | 0.25 |
| O | 48e | 0.39037(2) | 0.15470(2) | 0.381852(19) |

Table 1: $In_2O_3$ Atomic coordinates determined by XRD. The occupation of all sites is 1.

#### 2.1.4 Resistivity Measurements

The resistivity of the $In_2O_3$ crystals was measured in an ambient atmosphere using the four-point-probe method (frequency range: 1 - 10 Hz) without any pretreatment of the sample. The resistivity at room temperature is (2 ± 0.5) × $10^5$ Ω cm.

### 2.2 Surface Characteristics

#### 2.2.1 Surface Quality

The sample surface preparation consisted of several sputter and annealing cycles in an UHV-chamber (base pressure < 5 × $10^{-10}$ mbar). A sputtering cycle included a ten-minute surface bombardment with 2 keV argon ions (current density ≈ $10^{-5}$ A/cm$^2$). During the annealing excursions the sample was kept at a temperature of 500°C for 10 minutes. Annealing to higher temperatures (600°C) caused an irreversible color change of the crystal to orange and was therefore avoided; this change is likely caused by the desorption of oxygen and the formation of additional oxygen vacancies.[19] After several sputter and annealing cycles the surface showed a sharp LEED pattern as shown in Figure 3 (a). These results compare well with the LEED results reported earlier by Morales and Diebold,[9] which were obtained on pure and tin-doped $In_2O_3$ (001) thin films grown with oxygen-plasma assisted MBE on Yttria-Stabilized Zirconia (YSZ). Similar to their results, a p4g glide plane symmetry is observed, which causes the extinction of the (0, n + 1/2) and (n + 1/2, 0) spots indicated by white boxes in Figure 3 (a).

Auger electron spectroscopy (AES), X-ray photoelectron spectroscopy (XPS) and LEIS with 1 keV helium ions and a scattering angle of 105° confirmed that the surface is free of contamination. LEIS, which is sensitive to strictly the top-most layer, showed an indium peak that is far larger than the oxygen peak (Figure 3 (b)).

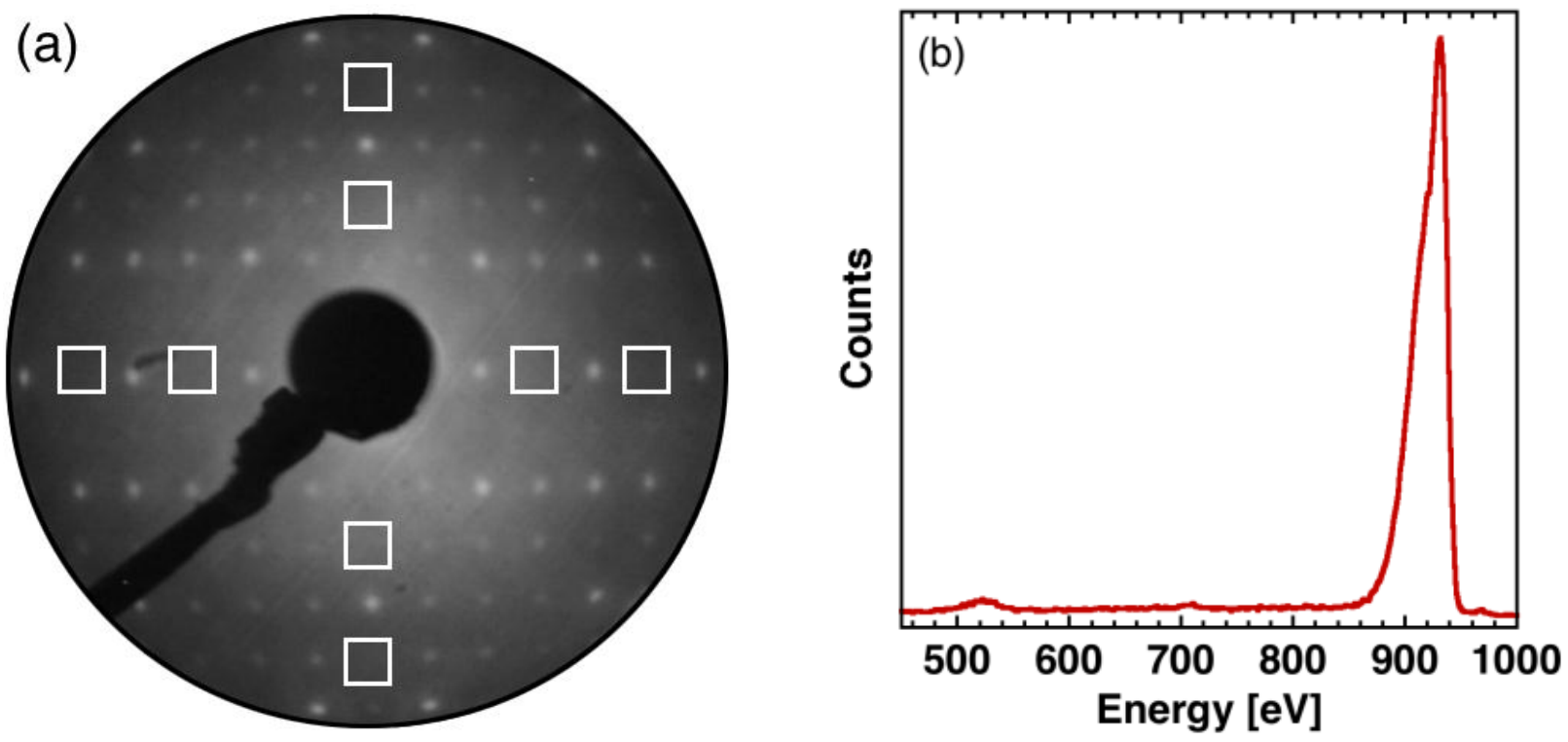

Figure 3 (a) LEED results of an $In_2O_3$(001) single crystal ($E_{Electron}$ = 100 eV) prepared by sputtering and annealing. Marked with squares are the (0, n + 1/2) and (n + 1/2, 0) spots that are missing due to the *p4g* glide plane symmetry. (b) $He^+$ LEIS spectrum that shows a far more pronounced indium peak ($E$ = 916.1 eV) compared to the oxygen peak ($E$ = 525 eV)

### 2.2.2 STM

Scanning Tunneling Microscopy was performed in UHV (base pressure < $5 \times 10^{-10}$ mbar) at room temperature with two different instruments (SPECS Aarhus design and a customized Omicron micro-STM) using electrochemically etched tungsten tips. Imaging was possible with both positive and negative bias with no apparent difference in the appearance of atomically-resolved images. A 100×100 $nm^2$ image of the sputtered/annealed surface is shown in Figure 4. It exhibits wide terraces with relatively regular step edges. The line profile shows that the terraces have a step height of 5 Å. After exposing the surface to activated oxygen with a plasma source, the same step height distribution was found, although the resulting surface appeared more complex in higher resolution images. The detailed results of these oxygen-exposure experiments will be reported in a future publication.

Figure 5 (a) shows a 20×15.75 $nm^2$ image with higher resolution. It is dominated by bright, atomic-sized protrusions, which at first seem to be arranged in a disordered way. Analysis with Fourier transforms, however, clearly reveals a square 3.6-Å lattice; this length equals a quarter of the diagonal of the $In_2O_3$ bulk unit cell. We have extracted this lattice by creating a binary image with only one nonzero pixel at the center of each atom and extracting the 3.6-Å periodicity in the Fourier domain. The resulting lattice positions are shown as dots in Figure 5 (b). It is obvious that most atoms are close to a lattice site of this regular 3.6-Å grid, but not directly on it. Only a few atoms are far from lattice sites - sometimes close to the center of a grid cell. The coverage of the protrusions in Figure 5 amounts to about two per unit cell, which is ¼ of the atomic density of the 3.6-Å lattice. The dark areas without protrusions are situated 2.5 Å below the protrusion level.

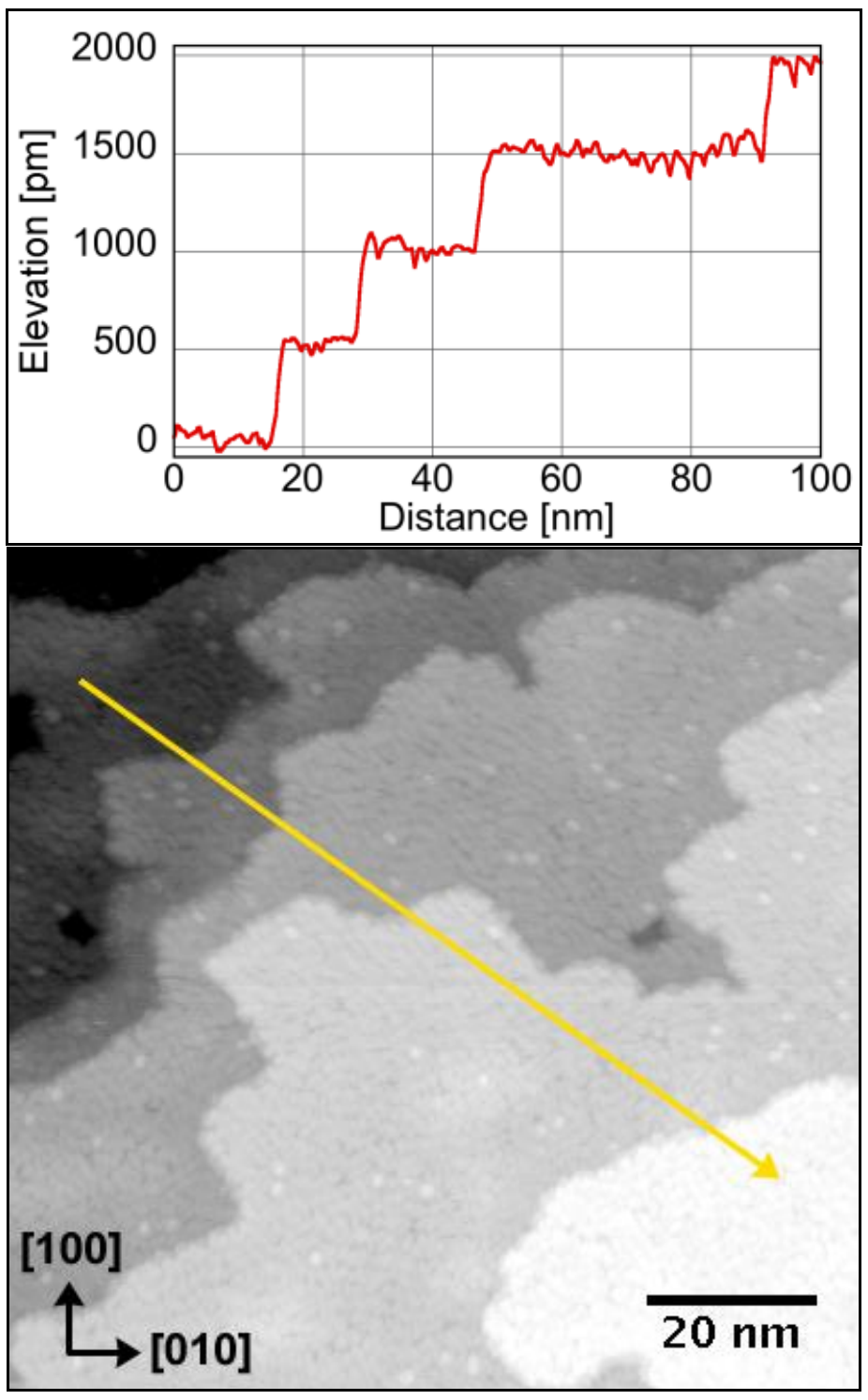

Figure 4: (Color online) STM image of a sputtered/annealed $In_2O_3(001)$ surface taken with a sample bias of $V_{Sample\ Bias}$ = -3 V and tunneling current $I_{Tunnel}$ = 0.225 nA. The arrow indicates the location of the line profile.

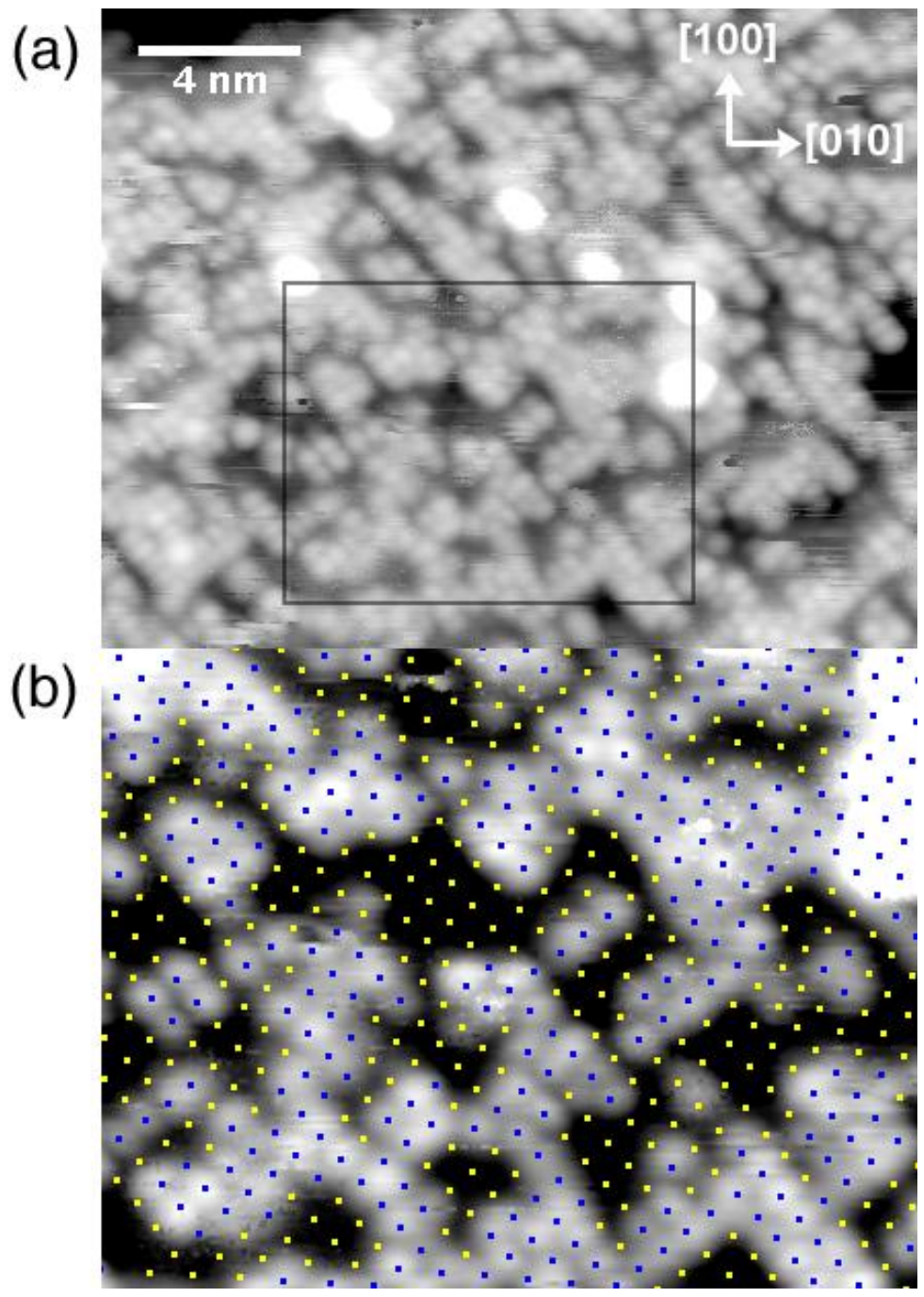

Figure 5 (Color online): (a) STM image with atomic resolution ($V_{\text{Sample bias}}$ = -2.6 V, $I_{\text{Tunnel}}$ = 0.17 nA). (b) This shows a magnification of the framed part above. The 3.6 Å lattice acquired from the Fourier transform is overlaid.

### 2.2.3 STS

STS was applied to measure the local density of states (LDOS) of the sample around $E_F$. A total of 400 spectroscopy points from a 20 × 20 grid in a 30 × 30 $nm^2$ area were acquired. A lock-in amplifier was used to directly obtain the differential conductance *dI/dV*, which is roughly proportional to the LDOS. The STS spectrum in Figure 6 shows the average over the 400 *dI/dV* curves. A strong state at $E_F$ is clearly visible.

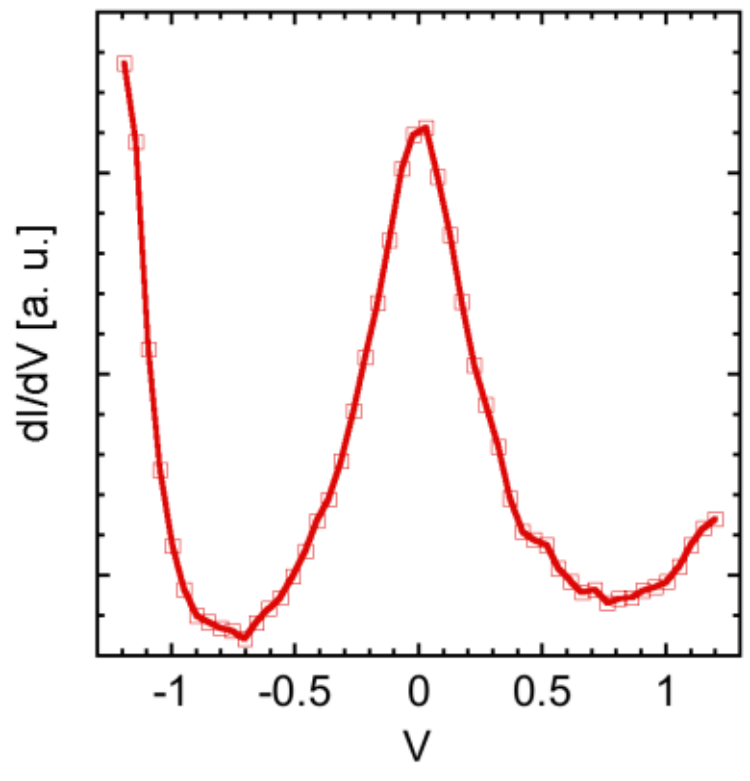

Figure 6 Scanning Tunneling Spectroscopy of a sputtered/annealed $In_2O_3$(001) surface.

### 2.2.4 PES

High-resolution photoelectron spectroscopy was carried out at the MAX II storage ring at MAX-lab in Lund, Sweden. The SX-700 monochromator at beamline I311[20] supplies photons in the energy range from 43 to 1500 eV. The spectroscopy end station is used for high-resolution XPS and NEXAFS measurements under UHV conditions (base pressure: < 3 $\times$ $10^{-10}$ mbar). The preparation chamber includes, among others, an argon ion sputter gun, LEED optics, mass spectrometer, and an atomic oxygen source (MBE-Komponenten OBS-40). A hemispherical energy analyzer (SCIENTA SES200) was used for the photoemission experiments. Measurements were performed at room temperature with the photon beam incident at 55° to the sample surface and normal emission.

For these measurements, the $In_2O_3$ crystal was mounted on a rhodium crystal, which allowed us to calibrate the Fermi energy by measuring the rhodium surface next to the $In_2O_3$ sample. It is defined as 0 eV binding energy for all of the spectra shown here. The valence band (VB) region was acquired with photon energies of 60, 85, 105 and 185 eV as shown in Figure 7 (b). Due to the small size of the $In_2O_3$ crystals, it turned out to be difficult to completely avoid rhodium peaks (the rhodium counts could be minimized to about 1% of the In $3d_{5/2}$ counts). The photon energy of 105 eV was chosen because the Rh 4d-levels have a cross-section minimum at this energy.[21]

Sample preparation was identical to that used for STM analysis. To oxidize the surface, the crystal was treated with atomic oxygen for 30 minutes at an oxygen pressure of 3 $\times$ $10^{-7}$ mbar and a flux of ~ 5 $\times$ $10^{14}$ oxygen atoms $cm^{-2}$ $s^{-1}$. While dosing, the sample was kept at a temperature of 200 °C or 400 °C.

Figure 7 (a) shows a comparison of the VB of the reduced and oxidized surfaces. The valence band maximum (VBM) is determined by extrapolating a linear fit of the leading edge of the valence band photoemission to the background level.[22] The VBM was determined using the data displayed in Figure 7 (a) and lies at 2.9 eV and 2.4 eV below $E_F$ for the reduced and the oxidized surfaces, respectively. A shift of the same magnitude can be seen for the In $3d_{5/2}$ core levels (0.52 ± 0.11 eV, Figure 8 (a)) and the O 1s core level (0.54 ± 0.11 eV, Figure 8 (b)). The In 4d semicore levels are shifted by a smaller amount (≈ 0.3 eV); probably they are influenced by the VB.

A closer look at the O 1s peak of the oxidized surface reveals a shoulder at the higher binding energy, shifted by 1.84 eV with respect to the main peak as shown in Figure 8 (c). Further noteworthy features in Figure 7 (a) are the pronounced gap states of the sputtered/annealed surface, which completely disappear for the oxidized surface.

Figure 7 (b) shows the VB of the reduced surface at different photon energies. It is apparent that the intensity of the gap states decreases with increasing photon energy, i.e., with decreasing surface sensitivity.

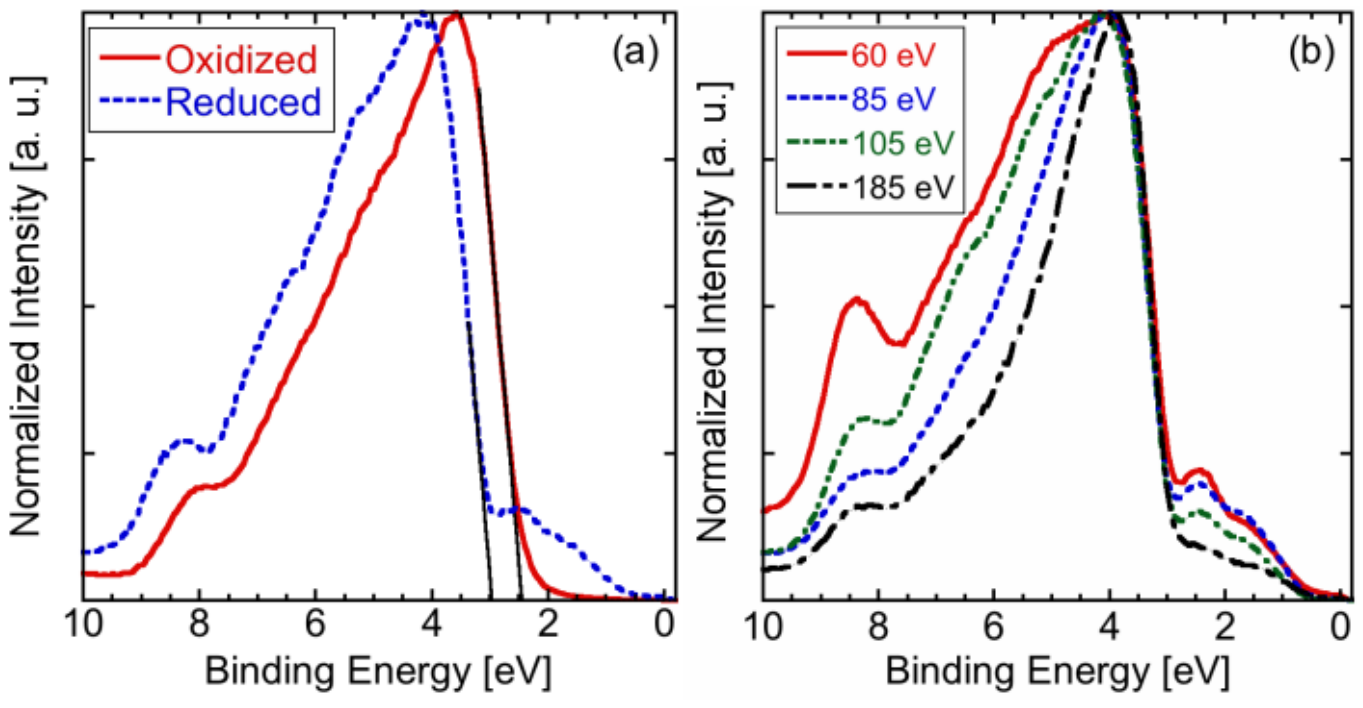

Figure 7 (Color online) (a) Normalized valence band spectra of the oxidized and sputtered/annealed $In_2O_3$(001) sample at 105 eV photon energy. Linear extrapolations for determining the leading edge of the VB are indicated. (b) VB of the reduced surface at different photon energies.

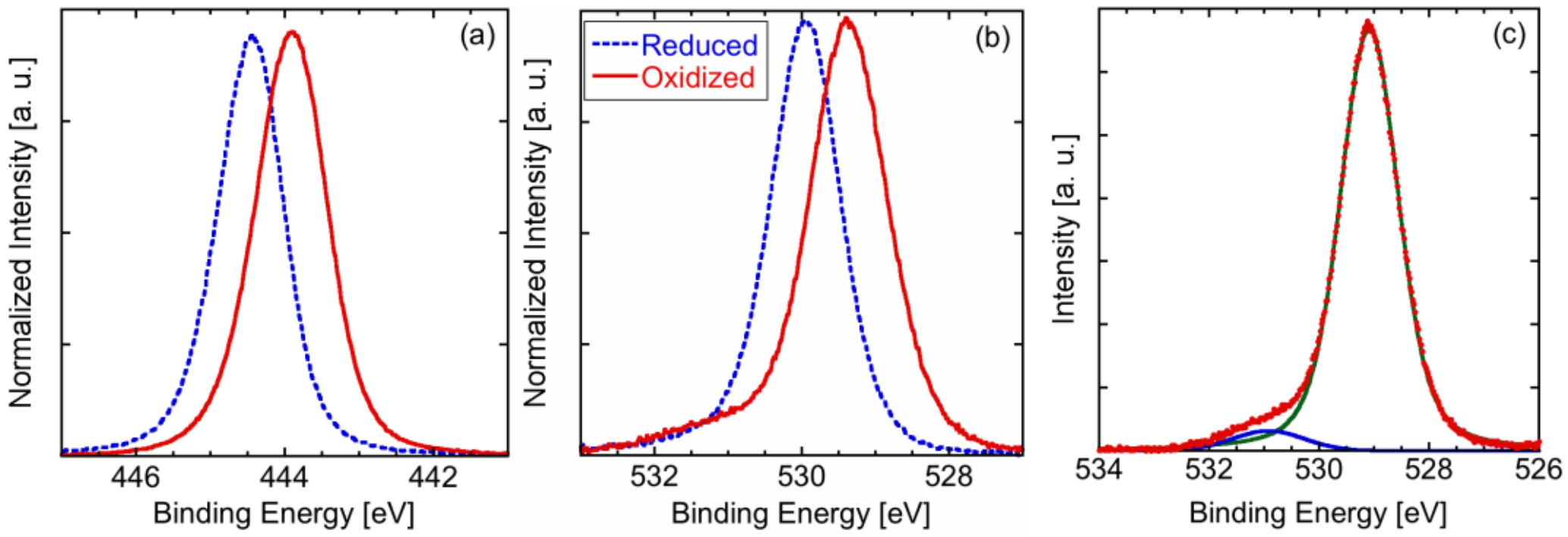

Figure 8 (Color online) PES of the (a) indium $3d_{5/2}$ and (b) O 1s core levels, taken with a photon energy of 610 eV. (c) Line fits of the O 1s peak of the oxidized surface at 610 eV (c).

# 3 Discussion

## *3.1 Surface Structure*

As shown in Sec. 2.2.2 the step edges of the $In_2O_3$(001) surface have a height of exclusively 5 Å. This has an important consequence if one considers the crystal structure of $In_2O_3$ (Figure 1): the distance between equivalent layers (e.g. from M- to M-layer) is roughly 5 Å. However, adjacent layers (i.e., from M- to D-layer) have a distance of about 2.5 Å. The observation of 5 Å-high steps implies that the large terraces have one particular termination, either M or D.

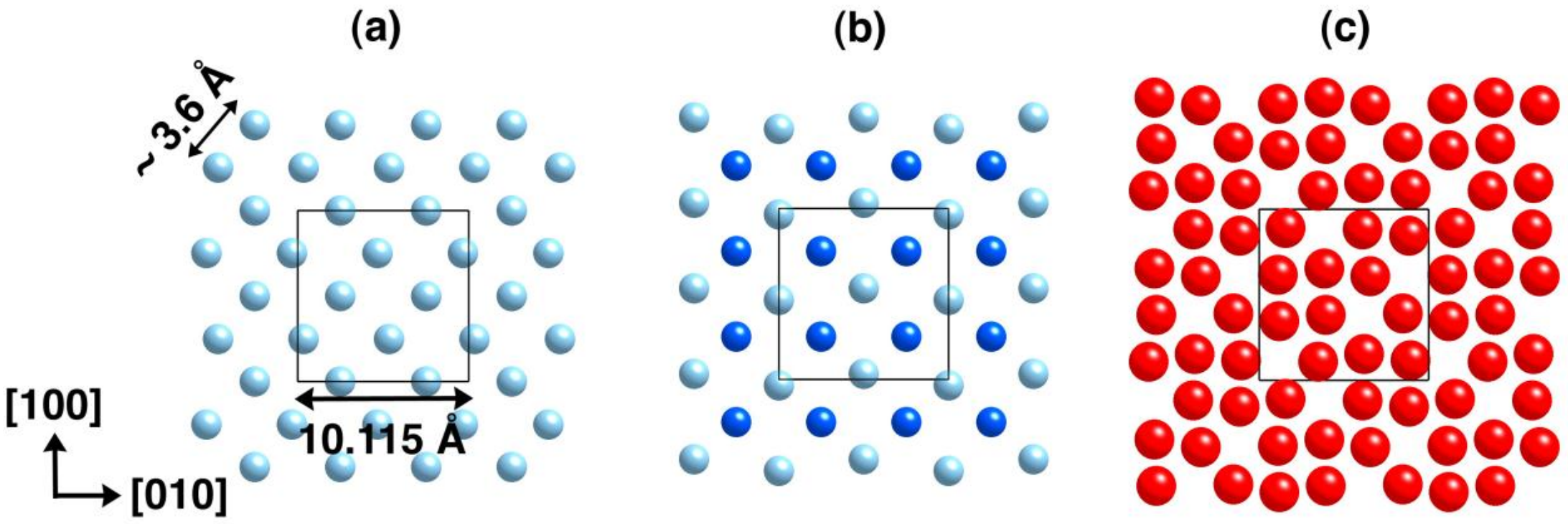

Figure 9 (Color online) Models of the $In_2O_3$(001) surfaces with (a) D-layer, (b) M-layer (b), and (c) O-layer termination. Shown are bulk-terminated layers with all the indium and oxygen atoms present. Note that these terminations represent polar surfaces; for stoichiometric surfaces only half of the atoms shown here are expected to be present.

In a recently published paper by Agoston and Albe,[17] possible terminations for low-index $In_2O_3$ surfaces were investigated in detail by DFT calculations. Both, stoichiometric (i.e., half of the indium/oxygen atoms missing from the bulk-terminated surfaces that are shown in Figure 9) and non-stoichiometric surfaces were considered as a function of the oxygen chemical potential. For the stoichiometric surfaces, these calculations show that the half-oxygen terminations are energetically preferable compared to the half-indium terminations, in agreement with previous DFT results from Walsh and Catlow.[1] The differences are very small, however. Due to the small energy differences between the different terminations, Agoston and Albe conclude that considerable disorder has to be expected and that even a coexistence of different terminations could be possible.

The surface stoichiometry depends on the oxygen chemical potential. Agoston and Albe found that different surface terminations become energetically preferable, when going from highly reducing to highly oxidizing conditions: (i) additional indium nucleates on top of the surface; (ii) 2 oxygen atoms per unit cell (M-layer), (iii) 4 oxygen atoms per unit cell (D-layer); (iv) stoichiometric termination with six surface oxygen atoms; (v) partial dimerized termination with two dimers (i.e. peroxo species) and four oxygen atoms; (vi) 6 dimers (M-layer). A dimerized surface would be in agreement with other DFT-calculations, which predict that surface oxygen atoms undergo a dimerization.[3,23]

For the sputtered/annealed surface, we can safely rule out a (partial) peroxide termination for the following reasons: (1) the preparation included annealing to 800 K in an oxygen background pressure of less than $10^{-12}$ mbar. This correspond to an oxygen chemical potential of approximately ~-2 eV assuming equilibrium with the residual gas (which is unlikely, however). This would lead to a metallic termination according to Ref. 17. An oxygen-rich termination can be ruled out because oxygen has been depleted by sputtering; (2) the top-layer consists predominantly of indium according to LEIS (Figure 3 (b)); (3) according to simulated STM images, oxygen dimers should appear dark;[17] (4) as suggested by Gassenbauer et al,[24] a surface peroxo species should show an O 1s core level with an additional structure at the higher BE side. While we, indeed, see such a peak for oxidized surface in Figure 8 (c), it is absent for the sputtered/annealed surface.

At first sight, the atom-sized features in Figure 5 (a) seem to be arranged in an unsystematic way. As shown in Figure 5 (b), most protrusions (atoms) are close to lattice sites of a 3.6 Å lattice, rotated by 45° with respect to the axes of the unit cell. The ball model in Figure 9 (c) shows that all oxygen atoms are situated close to lattice sites of a square lattice with $a/4 \approx 2.5$ Å, aligned with the axes of the unit cells. We have examined whether the protrusions visible in the STM image (Figure 5) could fit to such a lattice, but we found no solution that is compatible with the fourfold symmetry of the surface. The arrangement of the indium atoms is a distorted 3.6 Å lattice rotated by 45°, for both, D- and M-terminations (see Figure 9 (a), (b)). This suggests that the protrusions visible in the STM image of the sputtered/annealed surface are indium atoms, but that this indium lattice is only partially occupied. The number of protrusions in the STM image is determined as 2 per unit cell. This corresponds to half the indium atoms expected for a stoichiometric In-terminated surface (or 1/4 of the density of the full indium D- or M-layer shown in Figure 9 (a) and (b)).

Comparing the atoms in Figure 5 with the simulated STM images for the reduced surface, the elongated shape of the atoms could be an indication that we are seeing the D-layer. The unresolved, dark areas in Figure 5, ~2.5 Å below the protrusions, would then be the M-layer. The surface would thus consist of an M-layer at the basis, with small, irregular patches of a D-layer on top. The deviations of the atomic positions from a square 3.6-Å lattice, as seen by STM, are larger than those in either the D- or the M-layer, however, they might be related to either a density of states that is not centered at the atoms or due to mutual repulsion of the positively charged indium atoms distorting the lattice.

The occurrence of small islands on a polar metal-terminated oxide surface is reminiscent of the ZnO(0001) case, where it was found that O-terminated step edges provide additional negative charge, compensating for the excess positive charge of the zinc.[25] We consider it possible that the same mechanism explains the occurrence of the small indium islands on the $In_2O_3$(001) surface.

Interestingly, the STM results presented here show quite a different appearance from the results obtained earlier by Morales and Diebold.[9] They investigated thin films of ITO (001) grown on YSZ and prepared with an oxygen plasma source, i.e., under highly oxidizing conditions. Protrusions arranged in a zig-zag pattern were observed and interpreted as oxygen dimers of the peroxide termination. Cross-shaped dark features were attributed to missing dimers. The samples investigated in Ref. 9 differ in various ways from the single crystals studied here. They have been grown in a highly oxidizing environment; they consisted of relatively thin epitaxial films that may have been influenced by strain induced by the YSZ substrate; and they were doped with tin. All three parameters are expected to affect the surface termination - although it is left to future investigations to determine to what extent.

The photoelectron spectrum of the O 1s CL of the oxidized surface (Figure 8 (c)) shows a shoulder attributed to the presence of peroxide species. Agoston and Albe[17] considered a 'full' peroxide termination (6 dimers/unit cell) and the 'partial' peroxide surface (2 dimers, 4 oxygen atoms). Assuming we probed three monolayers with PES (due to the photoelectron escape length at a kinetic energy of ~100 eV), the expected ratio of oxygen in peroxide versus oxygen in bulk-like positions should be ~1/3 and ~1/10 for the full and partial peroxide termination, respectively. The intensity ratio of the shoulder to the main O*1s* peak in Figure 8 (c) amounts to 5.3%. This indicates that the oxidized crystal has a considerably lower peroxide coverage than the full peroxide termination.

## 3.2 Electronic Structure

### 3.2.1 Bulk Properties

We will initially discuss the results from the four-point-probe measurements. The accuracy of the resistivity values is limited by the fact that no distinction between bulk and surface conductivity is possible. Also surface states may play a role in the surface conductivity. This makes it difficult to determine the resistivity accurately, but they do give an order of magnitude that will be sufficient for the following considerations.

The carrier concentration $n$ can be calculated from the conductivity $\sigma$ and the carrier mobility $\mu$ using the relation:[26]

$$n = \frac{\sigma}{e\mu}$$

The conductivity $\sigma = (5 \pm 1.3) \times 10^{-6}\ \Omega^{-1}\ \mathrm{cm}^{-1}$ was calculated from the resistivity value from the four-point-probe measurement. The carrier mobility value $\mu = 32\ \mathrm{cm}^2\ \mathrm{V}^{-1}\ \mathrm{s}^{-1}$ was obtained from Ref. 4. The carrier concentration then results in $(1 \pm 0.2) \times 10^{12}\ \mathrm{cm}^{-3}$, which is about seven orders of magnitude smaller than the lowest value reported previously in the literature (see Ref. 4). This means our crystals have fewer impurities than the ones investigated in previous publications. This is emphasized by the fact that our crystals are yellow compared to the often reported green or black coloration of $In_2O_3$.[27] A possible reason for this extremely low carrier concentration might be the Pb and Mg residues in the crystals; they could compensate the natural n-type doping by a p-type doping.

Nevertheless, the carrier concentration value is probably still assessed to be too high to represent the intrinsic value, and it gives an upper limit for two reasons: First, water or surface states could have led to an increase of the conductivity at the surface. Second, the mobility of our crystals could be higher compared to the value from the literature because of the lower impurity scattering in our crystals.

The Debye screening length $\lambda_D$ of the crystal was estimated using the relation[28]

$$\lambda_D = \left( \frac{\varepsilon_0 \varepsilon_r k_b T}{n e^2} \right)^{1/2},$$

where $\varepsilon_0$ is the vacuum permittivity, $\varepsilon_r$ the relative permittivity, $k_b$ the Boltzmann constant, $T$ the temperature, $n$ the carrier concentration, and $e$ the elementary charge. For $\varepsilon_r$ the value 4 was used.[29] The resulting Debye length has a rather large value of 2.4 μm due to the low carrier concentration.

The bulk Fermi energy can be estimated from the bulk carrier concentration:[26]

$$E_F - E_C = \ln\left[ \frac{n}{2} \left( \frac{m^* k_b T}{2\pi\hbar^2} \right)^{-\frac{3}{2}} \right] k_b T,$$

where $E_F$ is the Fermi energy, $E_C$ the conduction band energy and $m^*$ the effective mass of electrons in the CB. The effective mass is 0.35 $m_e$.[30] The expected position of the Fermi level in the bulk is, therefore, 0.4 eV below the CBM at room temperature

### 3.2.2 Surface States

In general, the VB of $In_2O_3$ is dominated by O 2p-derived states with a lower VB peak from In 5s.[1,17,23,31] The CB is dominated by the In 5s states. There are two interesting effects regarding the electronic structure, which were observed by PES and STS - namely gap states and a rigid shift in the core levels. Both depend on the sample preparation conditions.

Gap states of pure $In_2O_3$ have been reported previously. Klein investigated the surface of reactively evaporated $In_2O_3$ films in situ by using synchrotron-excited photoemission.[32] He observed gap states depending on the oxygen pressures while preparing the films. He describes an increasing intensity of the gap states with decreasing pressure, which goes along with an increase of the optical absorption. He concluded that these gap states have their origin in different stoichiometries.

Agoston and Albe calculated the electronic structure of the reduced, stoichiometric, and the full peroxide termination.[17] The reduced surface shows In 5s-derived, half-filled metallic gap states which should have an acceptor character and, therefore, cause an upward band bending. For the stoichiometric surface, they predict – similar to Walsh and Catlow[1] - a split-off feature from the O 2p states close to the VBM, which is caused by undercoordinated oxygen atoms. For the surface under oxidizing conditions, peroxide surface states appear in the band gap coming from the anti-bonding p−π orbitals of the dimers. Other distinctive features for this termination are the bonding p−σ orbitals at the bottom of the VB and a ~5 eV splitting of the O 2s states of the dimers.

As shown in Figure 7 (a), the reduced surface shows low-lying gap states at ~2.4 eV and ~1.8 eV. These may be the In 5s states of an indium terminated surface calculated by Agoston and Albe, but in contrast to their suggestion, the states are completely below $E_F$ (filled).

Figure 7 (b) shows the VB of the reduced surface at different photon energies. The gap states apparently decrease with increasing energy. This indicates that the gap states are located at the surface. Another interesting feature of Figure 7 (b) is the decrease of the VB peak at ~8.2 eV with increasing photon energy. According to the electronic density of states (EDOS) calculations this peak should mainly arise from the In 5s states. Therefore, it represents another argument for an indium termination.

As $E_F$ at the surface is very close to the CBM (according to PES) the state measured only with STS at $E_F$ (Figure 6) could be either a surface state or the edge of the CB. We believe that it cannot be the CB edge alone because in this case we would expect a higher slope. Unfortunately, this surface state could not be confirmed by PES. There are two plausible explanations for this: First, STS is only sensitive for surface states at $\bar{\Gamma}(k_{\parallel} = 0)$. The PES analyzer had a rather wide acceptance angle of $\theta = 12°$. The measurable parallel component with this set up can be determined by:

$$k_{\parallel} = \sqrt{\frac{2m^{*}E_{Kin}}{\hbar^{2}}}\sin\frac{\theta}{2}.$$

This leads to the result that areas of $k_{\parallel}^{2}\pi = 0.95 Å^{-2}$ (photon energy: 105 eV) or $1.67 Å^{-2}$ (photon energy 185 eV) are covered. Therefore, the analyzer collects electrons from an area larger than the first Brillouin zone (BZ): $\left(\frac{2\pi}{a}\right)^{2} = 0.39 Å^{-2}$. A state at $\bar{\Gamma}(k_{\parallel} = 0)$ might not be visible if it comprises only a small fraction of the first BZ. Assuming that the state observed by STS is located close to $E_F$, has a parabolic dispersion, and energy of $\Delta E$ from the bottom of the band to $E_F$, we can estimate the fraction of the BZ occupied by this state:

$$k_{\|} = \sqrt{\frac{2m^{*}E_{Kin}\Delta E}{\hbar^{2}}} .$$

With an effective mass equal to that of the $In_2O_3$ CB ($m^{*} = 0.35 m_e$)[30] and assuming $\Delta E = 2k_b T$ at 300 K, the area includes only ~1 % of the area measured by the PES analyzer.

The second argument is related to the cross sections: The results from Agoston and Albe suggest that the surface state probably arises from In 5s states.[17] According to the literature, the In 5s photoionization cross sections should be rather small compared to the ones for O 2p (~6.7% at 105 eV photon energy).[21] Even at 185 eV – where the In 5s cross section is about 25% of the O 2p cross section – the surface state could not be observed.

The oxidized surface does not exhibit gap states in PES contrary to what we would have expected from the DFT calculations:[17] neither the predicted O 2s splitting and the orbital at the bottom of the VB (full peroxide termination) nor the gap states for the undercoordinated oxygen (partly peroxide surface) could be observed

### 3.2.3 Band Bending

A more-or-less rigid shift of the VB (Figure 7 (a)) and all core levels (Figure 8 (a) and (b)) is usually explained by band bending. Surface states in semiconductors result from breaking of the translational symmetry of the bulk. They can be either donor-like (usually closer to the VBM) or acceptor-like (usually closer to the CBM). The energy at which the gap state changes its character from predominantly donor-like to acceptor-like is the charge neutrality level (CNL) of the semiconductor.[33] Depending on their position relative to the Fermi level, they can be either neutral (occupied donor-like or unoccupied acceptor-like states) - or charged positively (unoccupied donor-like states) or negatively (occupied acceptor-like states).[33] Charged surface states cause a rearrangement of carriers close to the surface in order to screen the surface charge. This causes the upward or downward bending of the CB and VB relative to the Fermi level.[4]

The large Debye length of 2.4 µm implies that the band bending reaches far into the bulk. This is in good agreement with the experiment, where no reduction of the band bending is observable when increasing the photon energy, i.e., lowering surface sensitivity of the PES measurement. However, the carrier concentration at the surface should be higher due to the band bending ($E_F$ closer to the CBM; see discussion below), which would decrease the screening length.

To determine the direction of the band bending, the bulk Fermi energy (0.4 eV below the CBM) has to be regarded. The VB and the core levels of the oxidized surface are shifted by approximately the same value of 0.5 eV to lower binding energies. Assuming a fundamental band gap of 2.93 eV[7], the Fermi level is located 2.5 eV above the VBM in the bulk. At the surface, the Fermi level is pinned at 2.9 eV (2.4 eV) for the reduced (oxidized) surface above the VBM, according to the results from PES. This means that the VB and the CB are bent downward by 0.4 eV for the reduced surface (Figure 10 (a)), and bent upward by 0.1 eV or less for the oxidized surface (Figure 10 (b)).

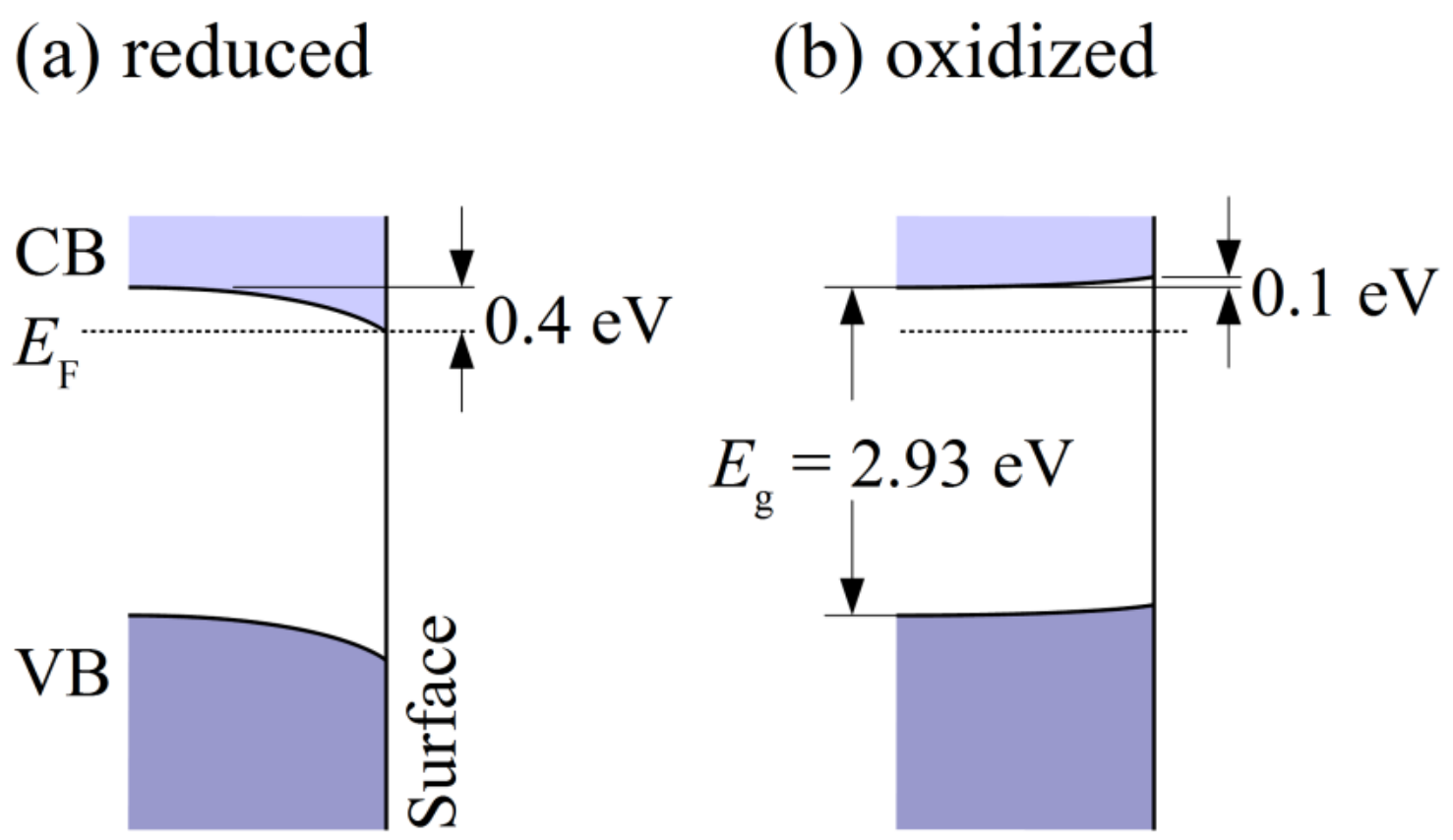

Figure 10: (Color online) Downward band bending of the reduced surface (a) and upwards band bending of the oxidized surface (b).

Band bending in $In_2O_3$ has been reported previously. King et al. described a downward band bending of 0.3 eV determined by XPS measurements and carrier statistics calculations for thin films grown by plasma-assisted molecular beam epitaxy (PAMBE) on YSZ.[4] This value was revised later to 0.45 ± 0.16 eV due to better knowledge of the band gap and an alternative method of analysis.[7] Their value is, nevertheless, in good agreement with our finding, where the VBM was determined by the common method of extrapolation of the leading edge of the VB (see Figure 7 (a)).

Klein observed that higher oxygen pressure during deposition leads to a lower Fermi level position in the band gap,[32] which is the same trend as for our results. Since he assumed a band gap of 3.6 eV, he explained the PES results with a surface depletion layer with an upward band bending of 1 eV.

For the reduced surface, the gap states visible close to the VBM are filled (occupied donor-like, neutral) and cannot cause band bending. The downward band bending can be explained by positively charged, unoccupied, donor-like surface states from the excess indium atoms. These lead to an accumulation of electrons at the surface that screen the positively charged surface state and cause the observed downward band bending. The excess indium is removed by oxidation, and the respective surface states and downward band bending disappear. The reason for the slight upward band bending, which indicates an electron depletion layer at the surface, is unclear. As it is very small, it might be only an artifact caused by uncertainties in determining the bulk Fermi level. Assuming a lower carrier concentration by a factor of 10 (as the (1 ± 0.2) × $10^{12}$ $cm^{-3}$ are an upper limit), the upward bend bending would almost disappear.

# 4 Conclusions

The bulk and surface properties of $In_2O_3$(001) single crystals have been investigated. The flux-grown samples permitted an accurate determination of the bulk lattice constant with XRD. Residues of Pb, Mg and Pt were found in the $In_2O_3$ crystals, which are characterized by a low carrier concentration of ≲ $10^{12}$ $cm^{-3}$.

Sputtering and annealing in UHV up to a maximum temperature of 500 °C (higher temperatures cause an irreversible orange discoloration) likely produces indium-terminated surfaces. These are characterized by low-lying gap states at ~2.4 eV and ~1.8 eV located at the surface. The position of the VBM at a binding energy of 2.9 eV implies a strong downward band-bending by 0.4 eV. A pronounced surface state was observed in STS. STM shows the presence of wide terraces that are separated by 5 Å high step edges. The terraces are covered with 2.5 Å high regions with bright, atomic-sized features in a square pattern aligned with the [110] direction. A possible interpretation is that the terraces consist of an M-layer with small “patches” of indium atoms of the D-layer.

When the surfaces are exposed to activated oxygen, the electronic structure changes. The low-lying band gap states disappear and the downwards band-bending is removed or even slightly reversed. A shoulder at the O 1s core level indicates the presence of peroxo species but probably not a completely dimerized termination.

# 5 Appendix

Details of the XRD refinement are given in the Table A.1. Atomic displacement parameters can be found in Table A.2; inter-atomic distances in Table A.3.

| | |
|---|---|
| Refinement on $F^2$ | 18 parameters |
| $R[F^2 > 2\sigma(F^2)] = 0.016$ | Weighting scheme based on measured s.u.'s 1/ $\sigma^2$ |
| $wR(F^2) = 0.017$ | $(\Delta/\sigma)max = 0.20$ |
| S = 3.61 | $\Delta\rho_{max} = 2.66$ e Å−3 |
| 910 reflections | $\Delta\rho_{min} = -2.77$ e Å−3 |

Table A.1: Refinement parameters.

| Site | Wyckoff Symbol | $U^{11}$ | $U^{22}$ | $U^{33}$ | $U^{12}$ | $U^{13}$ | $U^{23}$ |
|---|---|---|---|---|---|---|---|
| In1 | 8b | 0.004290(9) | 0.004290(9) | 0.004290(9) | 0.000580(5) | 0.000580(5) | 0.000580(5) |
| In2 | 24d | 0.003971(11) | 0.004030(10) | 0.004162(10) | 0 | 0 | 0.000542(5) |
| O | 48e | 0.00615(5) | 0.00592(5) | 0.00511(5) | -0.00046(4) | -0.00096(4) | -0.00027(4) |

Table A.2: Atomic displacement components.

| | | | |
|---|---|---|---|
| In1—O1$^{i}$ | 2.1734 (2) | In2—O1 | 3.8167 (3) |
| In1—O1$^{ii}$ | 2.1734 (2) | In2—O1$^{iii}$ | 3.8167 (3) |
| In1—O1 | 2.1734 (2) | In2—In2$^{xii}$ | 3.8392 (1) |
| In1—O1$^{iii}$ | 2.1734 (2) | In2—In2 | 3.8392 (1) |
| In1—O1$^{iv}$ | 2.1734 (2) | In2—In2 | 3.8392 (1) |
| In1—O1$^{v}$ | 2.1734 (2) | In2—In2 | 3.8392 (1) |
| In1—In2$^{vi}$ | 3.3445 (1) | In2—O1 | 4.0096 (3) |
| In1—In2$^{vii}$ | 3.3445 (1) | In2—O1 | 4.0096 (3) |
| In1—In2$^{viii}$ | 3.3445 (1) | In2—O1$^{i}$ | 4.1535 (3) |
| In1—In2$^{i}$ | 3.3445 (1) | In2—O1$^{xiii}$ | 4.1535 (3) |
| In1—In2$^{ii}$ | 3.3445 (1) | In2—O1 | 4.2583 (3) |
| In1—In2 | 3.3445 (1) | In2—O1 | 4.2583 (3) |
| In1—In2 | 3.8241 (1) | In2—O1 | 4.2766 (3) |
| In1—In2 | 3.8241 (1) | In2—O1 | 4.2766 (3) |
| In1—In2 | 3.8241 (1) | In2—O1 | 4.2912 (3) |
| In1—In2 | 3.8241 (1) | In2—O1 | 4.2912 (3) |
| In1—In2 | 3.8241 (1) | In2—O1$^{vi}$ | 4.3265 (3) |
| In1—In2 | 3.8241 (1) | In2—O1 | 4.3265 (3) |
| In1—O1 | 3.9926 (3) | In2—O1$^{vii}$ | 4.5346 (3) |
| In1—O1 | 3.9926 (3) | In2—O1$^{xiv}$ | 4.5346 (3) |
| In1—O1 | 3.9926 (3) | In2—O1$^{xv}$ | 4.7559 (3) |
| In1—O1 | 3.9926 (3) | In2—O1$^{xvi}$ | 4.7559 (3) |
| In1—O1 | 3.9926 (3) | In2—O1 | 4.9599 (3) |
| In1—O1 | 3.9926 (3) | In2—O1 | 4.9599 (3) |
| In1—O1 | 4.1002 (3) | O1—O1 | 2.7986 (3) |
| In1—O1 | 4.1002 (3) | O1—O1 | 2.7986 (3) |
| In1—O1 | 4.1002 (3) | O1—O1$^{xvii}$ | 2.8154 (3) |
| In1—O1 | 4.1002 (3) | O1—O1$^{vii}$ | 2.8154 (3) |
| In1—O1 | 4.1002 (3) | O1—O1 | 2.9387 (3) |
| In1—O1 | 4.1002 (3) | O1—O1$^{i}$ | 3.2856 (3) |
| In1—O1$^{vi}$ | 4.5334 (3) | O1—O1 | 3.2856 (3) |
| In1—O1$^{vii}$ | 4.5334 (3) | O1—O1 | 3.3119 (3) |
| In1—O1$^{viii}$ | 4.5334 (3) | O1—O1 | 3.3119 (3) |
| In1—O1$^{ix}$ | 4.5334 (3) | O1—O1 | 3.5488 (3) |
| In1—O1$^{x}$ | 4.5334 (3) | O1—O1 | 3.5488 (3) |
| In1—O1$^{xi}$ | 4.5334 (3) | O1—O1 | 3.7116 (3) |
| In2—O1 | 2.1249 (2) | O1—O1 | 4.1121 (3) |
| In2—O1 | 2.1249 (2) | O1—O1$^{vi}$ | 4.3137 (3) |
| In2—O1 | 2.1952 (2) | O1—O1 | 4.3137 (3) |
| In2—O1$^{xi}$ | 2.1952 (2) | O1—O1$^{xv}$ | 4.3468 (3) |
| In2—O1 | 2.2236 (2) | O1—O1 | 4.4732 (3) |
| In2—O1$^{iv}$ | 2.2236 (2) | O1—O1$^{iv}$ | 4.4732 (3) |
| In2—In2 | 3.3617 (1) | O1—O1 | 4.5196 (3) |
| In2—In2$^{vi}$ | 3.3617 (1) | O1—O1 | 4.6173 (4) |
| In2—In2 | 3.3617 (1) | O1—O1$^{ii}$ | 4.6173 (4) |
| In2—In2$^{vii}$ | 3.3617 (1) | | |

Table A.3: Interatomic distances.

Symmetry codes: (i) *z*, *x*, *y*; (ii) *y*, *z*, *x*; (iii) −*z*+1/2, −*x*+1/2, −*y*+1/2; (iv) −*y*+1/2, −*z*+1/2, −*x*+1/2; (v) −*x*+1/2, −*y*+1/2, −*z*+1/2; (vi) *z*, −*x*+1/2, *y*+1/2; (vii) *y*+1/2, *z*, −*x*+1/2; (viii) −*x*+1/2, *y*+1/2, *z*; (ix) −*z*+1/2, *x*, −*y*; (x) −*y*, −*z*+1/2, *x*; (xi) *x*, −*y*, −*z*+1/2; (xii) *z*+1/2, −*x*+1/2, −*y*; (xiii) *z*, −*x*, −*y*+1/2; (xiv) *y*+1/2, −*z*, *x*; (xv) *x*+1/2, *y*, −*z*+1/2; (xvi) *x*+1/2, −*y*, *z*; (xvii) *z*+1/2, *x*, −*y*+1/2.

# 6 Acknowledgments

This work was supported in part by the Austrian Science Fund (FWF; project F45) and the U.S. Department of Energy. Research at the Oak Ridge National Laboratory for one author (LAB) is sponsored by the U.S. Department of Energy, Basic Energy Sciences, Materials Sciences and Engineering Division. Useful discussions with Péter Ágoston und Karsten Albe are gratefully acknowledged.